\documentclass[sigconf,screen]{acmart}

\usepackage[utf8]{inputenc}

\usepackage{color}

\definecolor{grayblue}{rgb}{0.2,0.29,0.79}
\definecolor{darkgreen}{rgb}{0.2,0.55,0.1}
\definecolor{violet}{rgb}{0.54,0.17,0.88}

\newcommand{\zhou}[1]
{
   {\noindent\color{green}\bf [#1]$_{\scriptscriptstyle\textit{zhou}}$}
}

\newcommand{\tian}[1]
{
   {\noindent\color{red}\bf [#1]$_{\scriptscriptstyle\textit{tian}}$}
}
\newcommand{\guo}[1]
{
   {\noindent\color{grayblue}\bf [#1]$_{\scriptscriptstyle\textit{guo}}$}
}
\newcommand{\fang}[1]
{
   {\noindent\color{blue}\bf [#1]$_{\scriptscriptstyle\textit{fang}}$}
}

\newcommand{\includeAuthorComments}[1]
{
   \ifthenelse{\equal{#1}{0}}
   {
      \renewcommand{\fang}[1]
      {
         {} 
      }
      \renewcommand{\guo}[1]
      {
         {} 
      }
     \renewcommand{\tian}[1]
      {
         {} 
      }
     \renewcommand{\zhou}[1]
      {
         {} 
      }
   }{}
}

\AtBeginDocument{%
  \providecommand\BibTeX{{%
    \normalfont B\kern-0.5em{\scshape i\kern-0.25em b}\kern-0.8em\TeX}}}

\graphicspath{{figs/}}

\usepackage{graphicx}
\usepackage{subfigure}
\usepackage{booktabs}
\usepackage{bigstrut,multirow,rotating,booktabs}
\usepackage{amsmath}
\usepackage{setspace} 

\usepackage{CJKutf8}
\usepackage{algorithm}
\usepackage{algorithmicx}
\usepackage{algpseudocode}
\usepackage{amsmath}
\floatname{algorithm}{Algorithm}
\usepackage{bbm}
\usepackage{makecell}
\usepackage{array}
\usepackage{enumitem}

\copyrightyear{2024}
\acmYear{2024}
\setcopyright{acmlicensed}\acmConference[ASE '24]{39th IEEE/ACM International Conference on Automated Software Engineering }{October 27-November 1, 2024}{Sacramento, CA, USA}
\acmBooktitle{39th IEEE/ACM International Conference on Automated Software Engineering (ASE '24), October 27-November 1, 2024, Sacramento, CA, USA}
\acmDOI{10.1145/3691620.3695037}
\acmISBN{979-8-4007-1248-7/24/10}

\begin{document}


\title{SoVAR: Building Generalizable Scenarios from Accident Reports for Autonomous Driving Testing}

\author{An Guo}
\email{guoan218@smail.nju.edu.cn}
\affiliation{
  \institution{State Key Laboratory for Novel Software Technology \\Nanjing University}
  \city{Nanjing 210023}
  \postcode{210023}
\country{China}
}
\orcid{0009-0005-8661-6133}

\author{Yuan Zhou}
\authornote{Corresponding authors.}
\email{yuanzhou@zstu.edu.cn}
\affiliation{
  \institution{School of Computer Science and Technology 
  \\Zhejiang Sci-Tech University}
  \city{Hangzhou 310018}
  \postcode{310018}
\country{China}
}
\orcid{0000-0002-1583-7570}

\author{Haoxiang Tian}
\email{tianhaoxiang20@otcaix.iscas.ac.cn}
\affiliation{
  \institution{College of Computing and Data Science\\ Nanyang Technological University}
\country{Singapore}
}
\orcid{0000-0001-9132-9319}

\author{Chunrong Fang}
\authornotemark[1]
\email{fangchunrong@nju.edu.cn}
\affiliation{
  \institution{State Key Laboratory for Novel Software Technology \\Nanjing University}
  \city{Nanjing 210023}
  \postcode{210023}
\country{China}
}
\orcid{0000-0002-9930-7111}

\author{Yunjian Sun}
\email{sunyunjian.syj@smail.nju.edu.cn}
\affiliation{
  \institution{State Key Laboratory for Novel Software Technology \\Nanjing University}
  \city{Nanjing 210023}
  \postcode{210023}
\country{China}
}
\orcid{0009-0005-2259-0413}

\author{Weisong Sun}
\email{weisong.sun@ntu.edu.sg}
\affiliation{
  \institution{College of Computing and Data Science\\Nanyang Technological University}
\country{Singapore}
}
\orcid{0000-0001-9236-8264}

\author{Xinyu Gao}
\email{xinyugao@smail.nju.edu.cn}
\affiliation{
  \institution{State Key Laboratory for Novel Software Technology \\Nanjing University}
  \city{Nanjing 210023}
  \postcode{210023}
\country{China}
}
\orcid{0009-0004-7135-1833}

\author{Anh Tuan Luu}
\email{anhtuan.luu@ntu.edu.sg}
\affiliation{
  \institution{College of Computing and Data Science\\Nanyang Technological University}
\country{Singapore}
}
\orcid{0000-0001-6062-207X}

\author{Yang Liu}
\email{yangliu@ntu.edu.sg}
\affiliation{
  \institution{College of Computing and Data Science\\Nanyang Technological University}
\country{Singapore}
}
\orcid{0000-0001-7300-9215}

\author{Zhenyu Chen}
\authornotemark[1]
\email{zychen@nju.edu.cn}
\affiliation{
  \institution{State Key Laboratory for Novel Software Technology \\Nanjing University}
  \city{Nanjing 210023}
  \postcode{210023}
\country{China}
}
\orcid{0000-0002-9592-7022}

\renewcommand{\shortauthors}{An Guo et al.}

\begin{abstract}

Autonomous driving systems (ADSs) have undergone remarkable development and are increasingly employed in safety-critical applications. However, recently reported data on fatal accidents involving ADSs suggests that the desired level of safety has not yet been fully achieved. Consequently, there is a growing need for more comprehensive and targeted testing approaches to ensure safe driving. 
Scenarios from real-world accident reports provide valuable resources for ADS testing, including critical scenarios and high-quality seeds.
However, existing scenario reconstruction methods from accident reports often exhibit limited accuracy in information extraction. Moreover, due to the diversity and complexity of road environments, matching current accident information with the simulation map data for reconstruction poses significant challenges.


In this paper, we design and implement SoVAR, a tool for automatically generating road-generalizable scenarios from accident reports. SoVAR utilizes well-designed prompts with linguistic patterns to guide the large language model~(LLM) in extracting accident information from textual data. Subsequently, it formulates and solves accident-related constraints in conjunction with the extracted accident information to generate accident trajectories. Finally, SoVAR reconstructs accident scenarios on various map structures and converts them into test scenarios to evaluate its capability to detect defects in industrial ADSs. We experiment with SoVAR, using the accident reports from the National Highway Traffic Safety Administration’s~(NHTSA) database to generate test scenarios for the industrial-grade ADS Apollo. The experimental findings demonstrate that SoVAR can effectively generate generalized accident scenarios across different road structures. Furthermore, the results confirm that SoVAR identified 5 distinct safety violation types that contributed to the crash of Baidu Apollo.


\end{abstract}



\maketitle

\section{Introduction} 

The advent of autonomous driving technology has ushered in a new era of transportation, promising increased safety, efficiency, and convenience~\cite{DBLP:journals/ftcgv/JanaiGBG20,DBLP:journals/access/YurtseverLCT20}. However, the occurrence of crashes involving autonomous driving vehicles, including those that have resulted in fatalities~\cite{crash-bicycle2, crash}, serves as evidence that autonomous driving is not currently as safe as it is marketed to be. In many instances, these crashes can be attributed to defective software, highlighting the urgent requirement for an enhanced approach to testing autonomous driving software~\cite{DBLP:conf/sigsoft/LouDZZ022}.

As one of the most critical quality assurance techniques, ADS testing has attracted the attention of both academia and industry~\cite{DBLP:conf/issre/LiLJTSHKI20,DBLP:conf/issta/GuoF022,DBLP:journals/tse/ZhongKR23}. The key testing techniques for ADSs can be classified into two categories: road testing and simulation testing. Road testing involves testing specific driving scenarios in closed autonomous vehicle proving grounds~\cite{DBLP:conf/sigsoft/LouDZZ022} or monitoring autonomous vehicles in real traffic~\cite{DBLP:conf/issre/ZhaoRFSS19}, but this approach requires a long period and extensive resources. Additionally, the acquisition of diverse critical test data that encompasses a wide range of realistic usage scenarios presents significant challenges for testers. Therefore, high-fidelity simulation testing methods have become imperative to the development and validation of autonomous driving technologies~\cite{DBLP:journals/tosem/TangZZZGLGLMXL23,guo2024semantic}. It conducts ADS testing in simulation platforms, such as LGSVL~\cite{DBLP:conf/itsc/RongSTLLMBUGMAK20} and CARLA~\cite{DBLP:conf/corl/DosovitskiyRCLK17}. One essential aspect of ADS simulation testing is the identification and construction of critical scenarios that may lead to accidents.

Software engineering researchers have proposed utilizing real-life accident cases to generate critical test scenarios~\cite{DBLP:conf/sigsoft/GambiHF19,DBLP:conf/aitest/GambiNAF22,erbsmehl2009simulation,DBLP:conf/icra/BashettyAF20,DBLP:conf/issta/Zhang023}. The main insight is that real car accidents present critical situations that pose significant challenges for self-driving cars. Consequently, recent research has mainly focused on scenario reconstruction from different driving data sources. This includes utilizing textual descriptions~\cite{DBLP:conf/sigsoft/GambiHF19}, accident sketches~\cite{DBLP:conf/aitest/GambiNAF22}, sensor data~\cite{erbsmehl2009simulation}, and video recordings~\cite{DBLP:conf/icra/BashettyAF20,DBLP:conf/issta/Zhang023} to generate essential scenarios. Compared to sensor data and video recordings of critical cases, textual descriptions of crashes are more accessible and abundant~\cite{DBLP:conf/aitest/GambiNAF22,hizal2009construction}. Accident report analysis approaches offer more comprehensive insights into collisions, including intricate details like weather, lighting, and road conditions, which are visually challenging to represent. However, the accuracy of information extracted through current scenario reconstruction methods based on accident reports is limited~\cite{DBLP:conf/sigsoft/GambiHF19}.
Furthermore, current methods can reproduce the accident scenarios only on the same road structure described in the accident report. 
It limits the application to simulation-based ADS testing because the road structures in the simulation may differ from those described in the accident reports. Reproducing accident scenarios on different roads is challenging and overwhelming~\cite{DBLP:journals/eaai/BaoHLL23,DBLP:journals/corr/abs-2206-09357}.




To bridge this gap, we propose an automatic and universal method, SoVAR, to reconstruct accident scenarios on different road structures from accident reports. 
The primary objective of SoVAR is to automatically reconstruct accident scenarios from accident reports on different roads within the simulation environment, which can then be used as initial seeds for ADS testing.
SoVAR first leverages carefully designed linguistic prompt patterns to guide the LLM in accurately extracting environment, road, and object movement information.
Based on the extracted information and the roads to reconstruct the scenarios, SoVAR formulates the constraints for the accident scenario.
By solving these constraints with constraint solvers, we can generate the accident trajectories of the accident participants and finally reconstruct the accident scenarios. 
Therefore, the generated scenarios can be executed on the required map. 
Furthermore, to use these scenarios for ADS testing, SoVAR transforms them into testing scenarios by identifying the ego vehicle (i.e., the vehicle controlled by the ADS under testing) and the NPC (non-player character) vehicles. The NPC vehicles will follow the computed trajectories, while the ego vehicle is controlled by the ADS instead of following a predetermined trajectory.



To evaluate the effectiveness of SoVAR, we reconstruct scenarios based on well-known NHTSA's accident reports~\cite{NHSTA} and the San Francisco map provided by the LGSVL simulator.
The generated scenarios are then used to test Baidu Apollo~\cite{apollo}.
Our experimental findings demonstrate that the accident information extraction approach of SoVAR outperforms all baseline methods. Furthermore, subsequent results highlight that SoVAR successfully generated road-generalizable scenarios across different road structures compared to the existing textual accident report reconstruction method. Additionally, our experiments reveal that converting reconstructed scenarios into test cases effectively identified 5 distinct types of safety violation behaviors in the industrial-grade software Apollo.

The main contributions of this paper are summarized as follows:


\begin{itemize}[leftmargin=*]

\item  \textbf{Method.} We propose an innovative method for automatically reconstructing crash scenarios from accident reports and testing ADS. Our method enhances LLM's ability to accurately extract textual information by designing linguistic patterns for prompts. We then generate driving trajectories that align with desired road structures by solving a set of driving constraints.

\item \textbf{Tool.} We implement the proposed approach into the automated testing tool, SoVAR. To support the open science community, we have made the source code available$\footnote{https://github.com/meng2180/SoVAR}$ and released the scenarios that led to the ADS collisions.

\item \textbf{Study.} We utilize SoVAR to assess the industry-grade ADS Baidu Apollo and find fatal collisions. The results demonstrate that, compared with state-of-the-art scenario reconstruction techniques, our method can extract accident information more accurately and generate generalized accident scenarios that can adapt to different road structures. Furthermore, our approach can be used for ADS testing and to identify various safety violations in ADS.

\end{itemize}

\section{Background}

\subsection{Motivation}

In the real world, the autonomous vehicle industry typically employs simulation-based testing, evaluation, and validation as the first step, followed by testing in controlled environments (such as closed roads and test sites), and finally progressing to testing in open road configurations~\cite{DBLP:conf/itsc/HuangWLZ16}. During the simulation testing phase, where autonomous vehicles encounter unknown scenarios, efficiently constructing challenging scenarios becomes a top priority. To address this need, several academic institutions~\cite{DBLP:journals/jits/ChelbiGS22,guo2024semantic,zhou2023specification,tang2023evoscenario} and companies~\cite{hadj2020evaluating,hadj2019full} have developed testing platforms that generate possible combinations of scenarios in an initial configuration, using only the necessary initial parameters and variables. Reconstructing scenarios from real accident reports aims to generate initial scenarios capable of detecting ADS defects. However, there are currently two main challenges in obtaining key information and performing scenario reconstruction:

\begin{figure}[]
	\centering
    \includegraphics[width=\linewidth, height=0.35\linewidth]{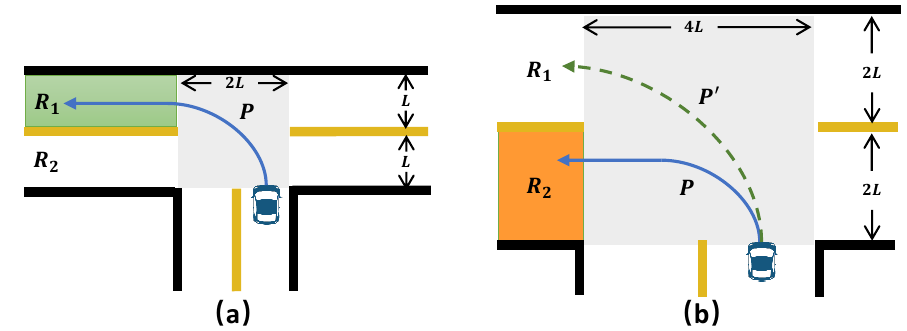}%
	\caption{A motivating example of reconstructing an accident scenario on different road structures.}
	\label{fig.2-1}
    \vspace{-10pt}
\end{figure}

\textbf{Challenge 1.} 
Accident reports encompass substantial information regarding the road environment and vehicle trajectories amidst traffic accidents~\cite{DBLP:conf/sigsoft/GambiHF19}. Nevertheless, the extraction of accident information encounters significant challenges. Off-the-shelf natural language processing (NLP) tools do not exhibit satisfactory performance due to the inclusion of traffic jargon and non-standard phrase structures in accident report~\cite{DBLP:conf/aitest/GambiNAF22}.

\textbf{Challenge 2.} 
ADS simulation testing is usually conducted in a simulator with different maps.
After extracting information from the accident report, reconstructing the accident scenario necessitates matching the road structures on the map to be tested in the simulator.
However, this task poses challenges for ADS engineers. On the one hand, the map covers all the key static properties of complex roads~\cite{liu2020high,ASAM}, including road type, road size (length and width), etc. It is extremely difficult to find roads that fully match the extracted information when reproducing the scenario within the map to be tested~\cite{DBLP:journals/eaai/BaoHLL23, DBLP:journals/corr/abs-2206-09357}.On the other hand, it is necessary to test multiple roads on the map because collisions often occur in similar road environments, thereby challenging the ability to adapt to a specific map. The current method~\cite{DBLP:conf/sigsoft/GambiHF19} does not include a mechanism to adapt the generated accident trajectory to the map. As shown in Figure~\ref{fig.2-1}, the accident that occurred in Figure~\ref{fig.2-1}(a) can also occur on the road in Figure~\ref{fig.2-1}(b). However, directly reconstructing the original trajectory on the right road cannot replay the accident. The vehicle turns left along the fixed trajectory $P$ to the road $R_1$ with a width of $L$. When the road width becomes $2L$, the vehicle turns left along the same trajectory $P$ to the reverse road $R_2$ with a width of $2L$, which is inconsistent with the expected trajectory P' and alters the semantics of the accident scenario (turn left to R1 and go straight -> turn left to R2 and retrograde). Therefore, planning a trajectory compatible with the simulation map during the scenario reconstruction process is necessary.



Based on the aforementioned challenges, we leverage LLMs to extract accident information and implement a suitable linguistic pattern prompting strategy to enhance extraction effectiveness. Additionally, we employ a constraint-solving strategy to generate trajectories of traffic participants, ensuring reproducibility on roads with varying lengths and widths, as well as different road types such as intersections and T-junctions.







\subsection{Large Language Models for Textual Understanding}
Large Language Models~\cite{kasneci2023chatgpt,DBLP:journals/corr/abs-2303-18223} (LLMs) have gained prominence in recent years, demonstrating comparable or superior performance to humans in various NLP tasks~\cite{DBLP:conf/acl/LaskarBRBJH23}, highlighting their ability to comprehend, generate, and interpret text. LLMs are sophisticated language models characterized by their massive parameter sizes and exceptional learning capabilities. Currently, LLMs are primarily employed through the utilization of prompts, which serve as concise cues or instructions that direct the model's output. This approach is commonly referred to as prompt-based learning.


In prompt-based learning, a pre-trained language model is optimized for various tasks through priming on natural language prompts~\cite{madotto2021few}. These prompts consist of text segments that are combined with an input and then used to generate an output for the given task. This approach has proven effective for few-shot and zero-shot learning in numerous general-domain tasks. Recent research has shown that large language models exhibit promising results in the few-shot setting, occasionally outperforming previous approaches that utilize fine-tuned models~\cite{DBLP:conf/nips/BrownMRSKDNSSAA20}. 
The ChatGPT~\cite{DBLP:journals/ieeejas/WuHLSLHT23} (Chat Generative Pre-trained Transformer) from OpenAI, has billions of parameters and is trained on a vast dataset encompassing textual understanding. Directly using GPT to understand complex text has shown mediocre performance; therefore, appropriate prompt patterns and rules need to be designed to achieve optimal performance in downstream tasks.

In this paper, we employ GPT-4~\cite{achiam2023gpt} to extract information regarding driving accidents. Furthermore, we design linguistic patterns for information extraction prompts to facilitate GPT-4's rapid adaptation to the task of extracting information from accident reports.

\section{Approach}

\begin{figure}[]
	\centering
    \includegraphics[width=\linewidth, height=0.55\linewidth]{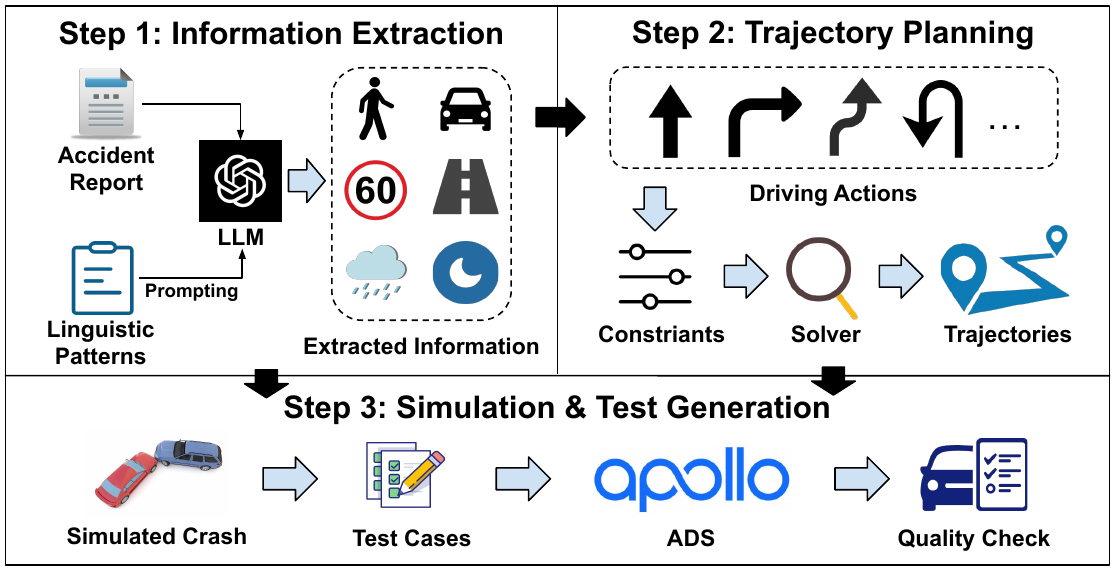}%
    \vspace{-5pt}
	\caption{The overview of SoVAR.}
	\label{fig3-1}
    \vspace{-10pt}
\end{figure}

In this section, we present the design and implementation of SoVAR, a tool devised for automatically reconstructing crash scenarios from accident reports and testing ADS. SoVAR consists of three steps to obtain simulation-based tests: information extraction, trajectory planning, and simulation and test generation, as illustrated in Figure~\ref{fig3-1}. 
According to the layer-based scenario definition \cite{bagschik2018ontology}, 
SoVAR initially abstracts the accident scenario into three layers: the road network and traffic guidance objects, the environmental conditions, and the dynamic objects. 
It then leverages an LLM to systematically extract information about accident-influencing factors from the accident report, organizing the information layer by layer. 
Following this, SoVAR establishes constraints on the pre-accident driving actions of traffic participants and employs a constraint solver to generate trajectories that comply with the specified constraints outlined in the accident report. Subsequently, the trajectories generated along with the extracted environment and road information are inputted into the driving simulator to reconstruct the car accident scenario. Finally, in the evaluation phase, SoVAR converts the generated simulation scenarios into test cases that include test oracles. These test cases are then inputted into the ADS. SoVAR checks whether the ADS under test successfully reaches its final expected position without encountering any crashes.

\begin{table*}[htbp]\footnotesize
  \centering
  \caption{Extracted accident information description and examples.}
  \vspace{-10pt}
    \begin{tabular}{c|c|c|c}
    \hline
    \textbf{Id} & \textbf{Attribute} & \textbf{Description} & \textbf{Examples} \bigstrut\\
    \hline
    \multicolumn{4}{c}{\textbf{Accident context- Environment Conditions information}} \bigstrut\\
    \hline
    1     & Weather & Weather conditions at the time of the accident & Weather="Cloudy" \bigstrut[t]\\
    2     & Lighting & Lighting conditions at the time of the accident & Lighting={"Dark"} \bigstrut[b]\\
    \hline
    \multicolumn{4}{c}{\textbf{Accident context- RoadNetwork and Traffic Guidance information}} \bigstrut\\
    \hline
    3     & LaneNum & Total number of lanes on the road where the accident occurred & LaneNum="4" \bigstrut[t]\\
    4     & CollisionLocation & Type of road on which the accident occurred & CollisionLocation="T-junction" \\
    5     & SpeedLimit & Speed limit for the road on which the car is travelling & SpeedLimit="60" \bigstrut[b]\\
    \hline
    \multicolumn{4}{c}{\textbf{Accident context- Dynamic Object information}} \bigstrut\\
    \hline
    6     & DrivingActions & Actions of traffic participants before the accident & DrivingActions=["P1:[turn right,...]","P2:[follow lane"...]",...] \bigstrut[t]\\
    7     & CrashType & Type of collision between the attacker and the victim at the time of the accident & CrashType="Rear-End" \\
    8     & DrivingDirections & Initial direction of travel for each traffic participant & DrivingDirections=["P1:westbound","P2:southbound"...] \\
    9     & RunningLanes & The initial lane position of each traffic participant & RunningLanes=["P1:1","P2:2"...] \\
    10    & ParticipantsNumber & Total number of traffic participants & ParticipantsNumber="2" \bigstrut[b]\\
    \hline
    \end{tabular}%
  \label{tab:extracted accident}%
\end{table*}%


\begin{table*}[htbp]\small
  \centering
  \caption{The example of linguistic patterns for information extraction prompts.}
  \vspace{-10pt}
    \begin{tabular}{c|c|m{11cm}}
    \hline
    \textbf{Id} & \textbf{Pattern type} & \multicolumn{1}{c}{\textbf{Sample of linguistic patterns/rules}} \bigstrut\\
    \hline
    1     &   Environment conditions information & You should help me extract environmental conditions. The answer includes <Weather> and <Lighting>. $\cdots$For the <Weather>, it means the weather conditions when the accident happened$\cdots$. $\cdots$If it's rainy when the accident happened, "rainy" should be added into the <Weather>$\cdots$.\bigstrut\\
    \hline
    2     & RoadNetwork and Traffic Guidance & You should help me extract roadnetwork and traffic guidance information. The answer includes <CollisionLocation>,<LaneNum>, and <SpeedLimit>. $\cdots$For the <CollisionLocation>, it means the type of road on which the accident occurred$\cdots$. $\cdots$If the accident happened near or at an intersection or intersecting roadway, the answer of <CollisionLocation> is "intersection"$\cdots$.\bigstrut\\
    \hline
    3     & Dynamic Object & You should help me extract dynamic object information. The answer includes <ParticipantsNumber>, <CrashType>, <DrivingDirections>, <RunningLanes> and <DrivingActions>. $\cdots$For the <DrivingActions>$\cdots$, if the car proceeds to do an intended action but does not do actually, such as intending to turn right, this intended action must not be added to the <DrivingActions>$\cdots$.\bigstrut\\
    \midrule
    \end{tabular}%
  \label{tab:2}%
  \vspace{-5pt}
\end{table*}%

\subsection{Information Extraction}\label{section:IE}

According to the official general standards~\cite{NHTSA-report} for recording accident reports issued by the U.S. Department of Transportation, accident reports usually encompass valuable information regarding the road environment and vehicle trajectories. To understand complex sentence structures and traffic terminology in these texts, SoVAR leverages the LLM to extract this information. During information extraction, SoVAR incrementally parses the narrative of the car crash and accumulates the information about weather, lighting, roads, and vehicles into a data structure that forms the abstract of a car crash. 
While LLM demonstrates excellence in information extraction tasks, its performance can be significantly influenced by the quality of its prompt. Specifically, we need to design an appropriate prompt that precisely describes what needs to be queried or requested to enhance the extraction accuracy and effectiveness of the LLM. Section~\ref{section:3.1.1} outlines the information that will be extracted, while Section~\ref{section:3.1.2} details how we organize this information into a format that LLM can better comprehend.


\subsubsection{\textbf{Layer-based Accident Abstract Representation.}} \label{section:3.1.1}



SoVAR only needs the descriptive text of the accident to extract information, without relying on additional data. Consistent with current work in scenario reconstruction~\cite{DBLP:conf/sigsoft/GambiHF19,DBLP:conf/issta/Zhang023}, SoVAR concentrates on the primary crash-contributing factors, namely lighting, weather, roads, and vehicle movements. It directly extracts information using LLM without introducing new information. To abstractly represent the accident information and organize it into a semantic structure that can be understood by the LLM, we present the extracted information in a hierarchical representation~\cite{scholtes20216}, divided into three layers: road, environment, and dynamic objects. Table~\ref{tab:extracted accident} shows detailed descriptions and examples of the extracted attributes. If information is missing from an accident report, it indicates that the missing details are not crucial to the accident. 
SoVAR demonstrates strong versatility because it automatically assigns default values when necessary.

\textbf{Environmental Conditions.} The environmental condition layer encompasses weather and light conditions. Weather conditions (Weather) include factors such as rain, fog, snow, and others. Light conditions (Lighting) pertain to the lighting conditions on the road, typically brighter during the day and darker at night. Additionally, lighting may be enhanced by the street lights.

\textbf{Road Network and Traffic Guidance.} This layer describes the road network and the traffic signs used for guidance on the road. Road represents the geographical context of a crash, including the type of road where the accident occurred (CollisionLocation) and the number of lanes on the relevant roads (LaneNum). Additionally, SoVAR extracts information about the speed limit (SpeedLimit) of the road to reconstruct the speed constraints applicable to the location of the accident.



\textbf{Dynamic Objects.} This layer contains information about the striker and victim involved in the crash and the moving actions that led to the crash. SoVAR extracts information regarding the number of traffic participants (ParticipantsNumber) involved in simultaneous collisions and identifies the type of collision. The collision type information (CrashType) specifies the angle at which the traffic participants collided, including three types of collisions: rear-end collision, frontal collision, and front-to-side collision. For each car involved in the accident, SoVAR extracts the status of each vehicle, including the initial running lane of the vehicle (RunningLanes), the initial running direction of the vehicle (DrivingDirections), and the vehicle's behavior before the crash (DrivingActions). Specifically, vehicle behaviors describe the regular and abnormal actions~\cite{najm2007pre} taken by the vehicle. Regular driving actions include U-turn, stop, drive into roads, vehicle cross, turn left, turn right, follow lane, and change lane. Abnormal driving actions include driving off the road and retrograde. Pedestrians involved in the accident are considered victims, and SoVAR constructs pedestrian cross and pedestrian walk actions for them.

\subsubsection{\textbf{Linguistic Patterns of Information Extraction Prompt.}}\label{section:3.1.2}

With the information to be extracted, we design linguistic patterns to generate prompts for the LLM. To design the patterns, each of the three annotators is tasked with writing a prompt sentence following the regular prompt template~\cite{DBLP:conf/www/ChenZXDYTHSC22,DBLP:conf/acl/GuHLH22}. Subsequently, we assess the impact of accident information extraction. Using these prompt sentences, the three annotators conduct card sorting~\cite{spencer2009card} and engage in discussions to derive linguistic patterns. As illustrated in Table~\ref{tab:2}, this process results in three linguistic patterns corresponding to the three sub-types of information outlined in Table~\ref{tab:extracted accident}. We show a simplified sample template for each linguistic pattern.

SoVAR extracts accident information layer by layer using the prompt patterns designed in Table~\ref{tab:2}. For each linguistic pattern, it first explains the meaning of each attribute to help LLM understand the extracted information. Additionally, the pattern includes heuristic rules to guide LLM in producing accurate results. For example, if a car intends to perform a certain action but a collision occurs before the action is executed, the intended driving action should not be extracted. 
Finally, we utilize few-shot learning to ensure the LLM's output conforms to our expected standards. Therefore, taking the linguistic patterns as references, the trajectory planning module can directly use the output to build trajectory constraints, which will be described in the next section.



\subsection{Trajectory Planning}\label{section:TP}

Accurately planning the trajectories of traffic accident participants that satisfy the extracted information and match the target roads is an essential process in accident scenario reconstruction~\cite{DBLP:conf/sigsoft/GambiHF19,DBLP:conf/aitest/GambiNAF22}. A trajectory is an ordered sequence of waypoints, i.e., positions and velocities a traffic participant must follow. For example, the $i^{t h}$ waypoint of the $b^{t h}$ action of vehicle $X$ can be represented as $X_i^b=(x, y, v)$, where $x, y, v$ are the $\mathrm{x}$-coordinate, $\mathrm{y}$ coordinate, and velocity, respectively. In addition, we introduce $ pos =(x, y)$ to represent the position of the waypoint in the floor plan.


SoVAR utilizes extracted driving actions and road information (including the roads from the accident report and the tested roads to construct and execute the scenarios) to simulate crashes and compute trajectories for simulated traffic participants. This approach first builds constraints on each action based on the road information and then generates trajectories by resolving the constraints.
Therefore, it can adapt to different roads on various maps. Algorithm~\ref{alg1} outlines the generation of waypoints. To facilitate the display in the algorithm, the extracted attributes are represented by letter abbreviations. For example, driving action is abbreviated as DA. The algorithm takes road information~$\mathbb{R}$ extracted from accident reports using LLM, extracted dynamic object information~$\mathbb{D}$, the given map~$\mathbb{MAP}$ in the simulator, and defined driving action constraints ~$\mathbb{C}$ as inputs. The algorithm initially parses the given map into a set of candidate roads, noting the road type (e.g., crossroads) for each road~(Line 3). It then iterates through the set of candidate roads, sequentially selecting lanes of varying lengths and widths~(Lines 4-5). If the current road type matches the type where the accident occurred and if the maximum number of lanes occupied by all traffic participants during their movement is less than or equal to the number of lanes on the selected road, the waypoint generation process begins~(Lines 6-7). Subsequently, following the adjustment of the driving direction and initial lane position of the traffic participant to match the selected road~(Lines 8-11), the constraint solver resolves constraints based on driving actions to generate waypoints~(Line 12). The following describes the design of driving constraints~$\mathbb{C}$ in detail.

\begin{CJK*}{UTF8}{gkai}
    \begin{algorithm}
        \caption{Waypoints generation process of trajectory planning}
        \begin{algorithmic}[1] 
            \Require {The extracted dynamic object information $\mathbb{D}$, the extracted road information $\mathbb{R}$, the given map $\mathbb{MAP}$, the set of driving constraints $\mathbb{C}$}
            \Ensure  Generated waypoints $genWPoints$
            \State$ CandidateRoad \leftarrow \emptyset$\;
            \State$ SelectRoad \leftarrow \emptyset$\;
            \State$ CandidateRoad=ParseMap(\mathbb{MAP}) $\;
            \For{$Cand$ in $CandidateRoad$}
                \State $SelectRoad=[Cand.len,Cand.wid,Cand.lnum,Cand.dir]$
                \If{$Cand.type==\mathbb{R}.CL$ }
                    \If{$ CalMaxLanes(\mathbb{R}.LN,\mathbb{D})	\leq Cand.lnum$}
                        \State$ Recal \leftarrow \emptyset$\;
                        \State $\mathbb{D}.RL.\mathbb{D}.DD=ConvertInfo(\mathbb{D},SelectRoad)$
                        \State$ Recal.append(\mathbb{D}.RL)$
                        \State$ Recal.append(\mathbb{D}.DD)$
                        \State$ GenWP =SolveTraj(\mathbb{C},\mathbb{D}.DA,Recal,SelectRoad) $\;
                    \EndIf
                \EndIf
            \EndFor
            \\
            \textbf{return} $GenWP$
        \end{algorithmic}
        \label{alg1}
    \end{algorithm}
\end{CJK*}


To make participants perform corresponding actions and drives into the crash site, we define a group of trajectory constraints for each action from five aspects. Then, SoVAR leverages a constraint solver to automatically generate trajectories for each participant. Besides, traffic participants must reach the collision location simultaneously while executing their actions. To accomplish this, SoVAR introduces a collision area $C_A$, which is automatically calculated by the system.


\begin{figure}[]
	\centering
    \includegraphics[width=\linewidth, height=0.7\linewidth]{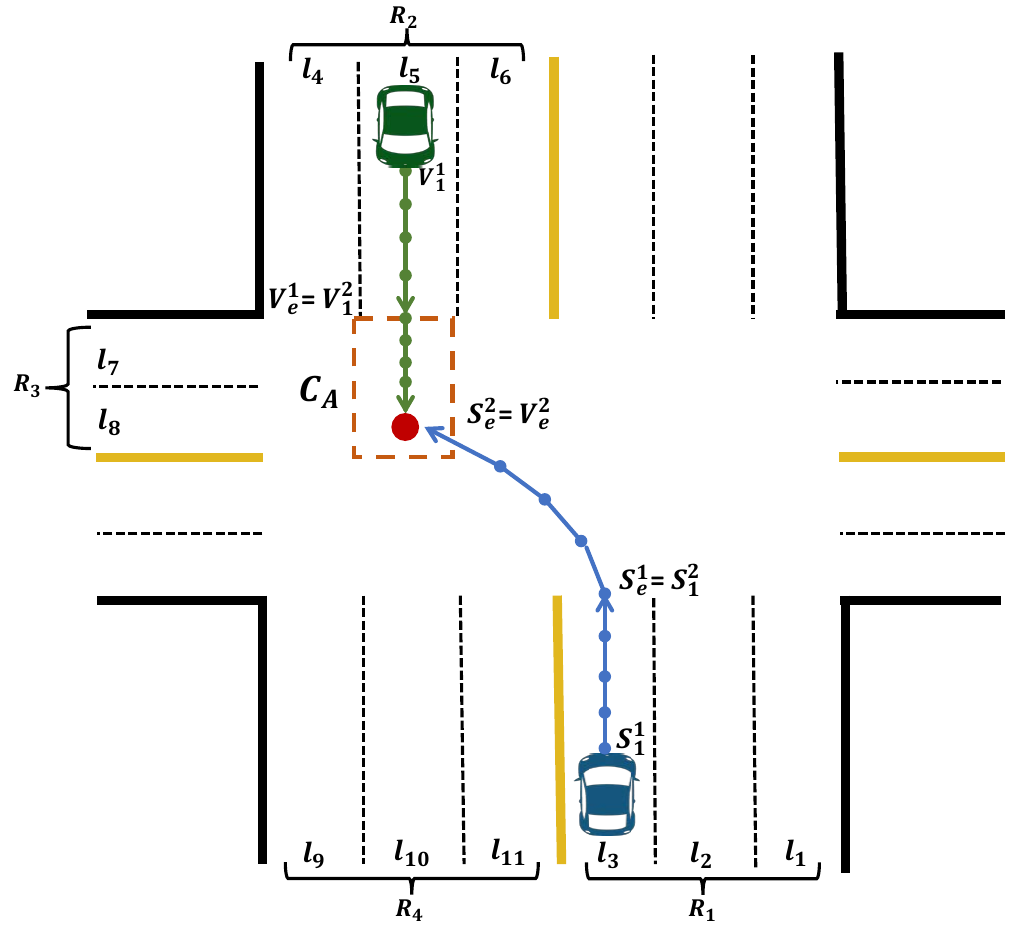}%
    \vspace{-10pt}
	\caption{An example of accident trajectory planning (NHTSA report~\#2006048103067)}
	\label{fig3-2}
    \vspace{-10pt}
\end{figure}


Due to space limitations, we have chosen an example to showcase these constraints. As shown in Figure~\ref{fig3-2}, the collision takes place at a regulated intersection. A blue vehicle is observed traveling northbound in lane $l_3$, making a left turn under a circular green signal, while a green vehicle is traveling south in lane $l_5$. During the blue vehicle's left turn, its front collided with the front of the green vehicle. The striker vehicle is depicted as following the lane and performing a left turn, while the victim vehicle is shown following the lane and performing a crossing maneuver. The blue vehicle may turn into lane $l_7$ or lane $l_8$. SoVAR calculates the intersection of the drivable areas between the green vehicle and the blue vehicle, determining the collision area $C_A$. Next, we introduce the constraint design of the driving actions involved in the example. The constraints for all accident driving actions can be found in~\url{https://github.com/meng2180/SoVAR/blob/main/constraints.pdf}.

\subsubsection{Constraints of Participant Basic Driving Actions} To generate trajectories of participants that make them perform basic driving actions, we define a group of trajectory specifications for each behavior from three aspects:1) initial position and destination, 2) position to perform the action, and 3) velocities of waypoints.



\textbf{Group 1: Constraints on initial position and destination.} 
There are four constraints in Group 1.
Equation \ref{eqn:g1-c1} restricts that the driving direction from $X_1^b$ to $X_e^b$ is not opposite to the direction of lane $l_m$. The calculation of $f d\left(w_i, w_j, l\right)$ is defined as $\left(w_j . x-w_i . x\right)(l_{ex}.x-l_{en}.x)>0 \wedge\left(w_j . y-w_i . y\right)(l_{ex}.y-l_{en}.y) >0$. Here, $l_{en}$ is the entrance point of lane $l$ and $l_{ex}$ is the exit point of the lane, which are both known after parsing the map. When $f d$ is true, it guarantees waypoint $w_j$ is ahead of $w_i$ in the direction of lane $l$.
Equation \ref{eqn:g1-c2} restricts the road position where the \textbf{\textit{Follow Lane}} behavior begins and ends. $X_1^b .pos$ and $X_e^b .pos$ are on the same lane $l_m$ of the road $R_i$.
Equations \ref{eqn:g1-c3} and \ref{eqn:g1-c4} limit the initial and the end road positions of the \textbf{\textit{Turn Left}} and \textbf{\textit{Vehicle Across}} actions:  
Equation \ref{eqn:g1-c3} is applied when the driving action avoids a collision, and $X_1^b .pos$ and $X_e^b .pos$ are on the lanes of different roads;
Equation \ref{eqn:g1-c4} is employed when the action leads to a collision, and the position of destination $w_e^X .pos$ is in $C_A$.
\begin{equation}\label{eqn:g1-c1}
    f d\left(X_1^b.\right.pos, \left.X_e^b . p o s, l_m\right)=1
\end{equation}
\begin{equation}\label{eqn:g1-c2}
   X_1^b . pos \in l_m, X_e^b . pos \in l_m, l_m \in R_i 
\end{equation}
\begin{equation}\label{eqn:g1-c3}
    X_1^b . pos \in l_m \wedge l_m \in R_i, X_e^b . pos  \in l_n \wedge l_n \in R_j, i \neq j 
\end{equation}
\begin{equation}\label{eqn:g1-c4}
    X_1^b . pos  \in l_m \wedge l_m \in R_i, X_e^b .  pos  \in C_A
\end{equation}

\begin{figure}[]
	\centering
    \includegraphics[width=\linewidth, height=0.5\linewidth]{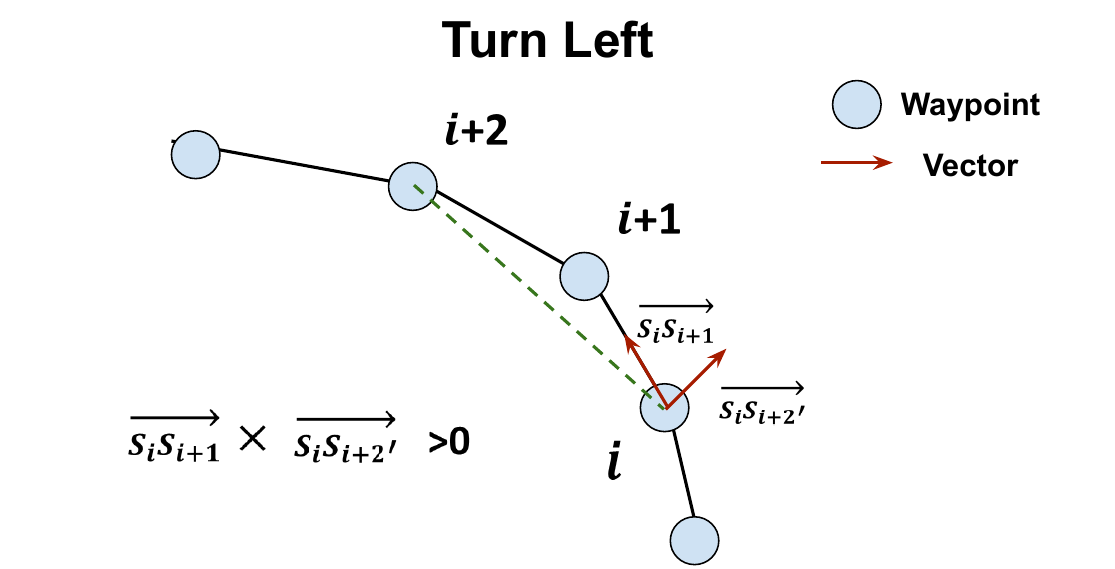}%
    \vspace{-10pt}
	\caption{A graphic explanation of Equation \ref{eqn:g2-c1} in Group 2.}
	\label{fig3-3}
\end{figure}

\textbf{Group 2: Constraints on positions to perform the action.}
In this group, Equation \ref{eqn:g2-c1} limits the position of the waypoint during the execution of the \textbf{\textit{Turn Left}} action, ensuring that the current waypoint is on the right side of the line connecting two adjacent waypoints. As shown in Figure~\ref{fig3-3}, $\overrightarrow{\mathrm{s_{i}s_{i+1}}}$ represents the direction vector composed of the position of the $i$-th waypoint and the position of the $(i+1)$-th waypoint. The vector $\overrightarrow{\mathrm{s_{i}s_{i+2}}^{\prime}}$ represents the right normal vector formed by the positions of the $i$-th and $(i+2)$-th waypoints.
Equations \ref{g2-c2} and \ref{g2-c3}
respectively impose limits on the positional relationship between the roads accessed through the execution of the \textbf{\textit{Turn Left}} behavior and the \textbf{\textit{Vehicle Across}} behavior. $k_1$ and $k_2$ represent the slopes of roads $l_m$ and $l_n$.


\begin{equation} \label{eqn:g2-c1}
    \forall i \in[1, e-1), \overrightarrow{\mathrm{s_{i}s_{i+1}}} \cdot \overrightarrow{\mathrm{s_{i}s_{i+2}}^{\prime}} >0
\end{equation}
\begin{equation}\label{g2-c2}
    \left\{
             \begin{array}{lr}
             k_1 * k 2=-1, l_{e n}^m \in l_m, l_{e x}^n, l_{e n}^n \in l_n  & \\
             k_1=\left(l_{e x}^m . y-l_{e n}^m . y\right) /\left(l_{e x}^m . x-l_{e n}^m . x\right) &  \\
             k_2=\left(l_{e x}^n . y-l_{e n}^n . y\right) /\left(l_{e x}^n . x-l_{e n}^n.x\right )
             \end{array}
\right.
\end{equation}

\begin{equation}\label{g2-c3}
  \left\{
             \begin{array}{lr}
             \arctan \left|\frac{k_1-k_2}{1+k_1 k_2}\right|<\frac{\pi}{2}, l_{e n}^m \in l_m, l_{e x}^n, l_{e n}^n \in l_n  & \\
             k_1=\left(l_{e x}^m . y-l_{e n}^m . y\right) /\left(l_{e x}^m . x-l_{e n}^m . x\right) &  \\
             k_2=\left(l_{e x}^n . y-l_{e n}^n . y\right) /\left(l_{e x}^n . x-l_{e n}^n.x\right )
             \end{array}
\right.  
\end{equation}

\textbf{Group 3: Constraints on the velocities of waypoints.} Finally, group 3 restricts the velocities of all intermediate waypoints. Specifically, Equation \ref{g3-c1} restricts velocities of waypoints from $X_1^b$ to $X_e^b$, and Equation \ref{g3-c2} restricts the limits of speeds during driving. $D X(i, j)$ and $D Y(i, j)$ respectively compute the distance that the vehicle traveled from the $i^{t h}$ waypoint to the $j^{t h}$ waypoint along $\mathrm{X}$ axis and $\mathrm{Y}$ axis. $DX(i,j)=\sum_{c=i}^{j-1}\left(\frac{v_c . x+v_{c+1} . x}{2} * \Delta t_{c}\right), D Y(i, j)=$ $\sum_{c=i}^{j-1}\left(\frac{v_c . y+v_{c+1} . y}{2} * \Delta t_{c}\right). v_c . x$ is the $c^{t h}$ waypoint's speed along $\mathrm{X}$ axis and $v_c . y$ is its speed along $\mathrm{Y}$ axis. $s p d_i$ is the speed of $v_i$. $s p d^{limit}$ is the speed limit for the driving road.
\begin{equation}\label{g3-c1}
    \left(X_1^b . x-X_e^b . x\right)=D X(1, e),\left(X_1^b . y-X_e^b . y\right)=D Y(1, e)
\end{equation}
\begin{equation}\label{g3-c2}
    \forall i \in(1, \mathrm{e}], 0<s p d_{i-1}^X=s p d_i^X \leq s p d^{limit}
\end{equation}





\subsubsection{Constraints between Multiple Driving Actions of Participants} To ensure that the striking and victim vehicles execute a sequence of driving actions and reach the accident site simultaneously, we define trajectory specifications from two aspects: 1) trajectory combinations of multiple basic driving actions, and 2) constraints on vehicle crashes.


\textbf{Group 4: Constraints on trajectory combinations of multiple basic driving actions.} The fourth group constrains the relationship between waypoints when multiple actions are connected. 
Equations \ref{g4-c1} and \ref{g4-c2}
represent the trajectory constraints for multiple actions exhibited by the striker and victim, respectively. The ending position of a traffic participant's current action should align with the starting position of the subsequent driving action. The striker performed $n$ actions, while the victim performed $m$ actions before the collision.
\begin{equation}\label{g4-c1}
    \forall p \in(1, \mathrm{n}], S_1^p . pos=S_{\mathrm{e}}^{p-1} . pos
\end{equation}
\begin{equation}\label{g4-c2}
    \forall q \in(1, \mathrm{m}], V_1^q .pos=V_{\mathrm{e}}^{q-1} . pos
\end{equation}





\textbf{Group 5: Constraints on vehicle crash.} The fifth group of constraints ensures that the collision information is consistent with the report description. 
Equation \ref{g5-c1} limits the occurrence of the accident collision to the location stated in the collision report, while 
Equation \ref{g5-c2} describes the striker and victim simultaneously arriving at the collision location.
\begin{equation}\label{g5-c1}
    S_{\mathrm{e}}^{n} . p o s=V_{\mathrm{e}}^{m} . p o s \wedge S_{\mathrm{e}}^{n} . p o s \in C_A \wedge V_{\mathrm{e}}^{m} . p o s \in C_A
\end{equation}
\begin{equation}\label{g5-c2}
    \sum_{i=1}^n S_{t_i}=\sum_{i=1}^m V_{t_i}
\end{equation}



\subsection{Simulation and Test Generation}\label{section:SG}

\textbf{Simulation generation.} SoVAR is not limited to a specific simulator; it requires a simulation environment that encompasses precise soft-body physics and authentic 3D textures representing roads, vehicles, weather, and lighting conditions. Additionally, the simulator must have the capability to control the virtual car using external software, as it serves as a fundamental requirement for conducting autonomous driving simulation testing~\cite{DBLP:conf/itsc/RongSTLLMBUGMAK20, DBLP:conf/corl/DosovitskiyRCLK17}. The simulator receives the waypoints generated by the trajectory planning module alongside the environment conditions information extracted by LLM and conducts a simulation to reconstruct the crash detailed in the accident report.

\begin{table*}[htbp]\scriptsize
  \setlength{\tabcolsep}{4pt}
  \centering
  \caption{Results of different methods for extracting information from accident reports.}
  \vspace{-10pt}
    \begin{tabular}{c|c|c|c|c|c|c|c|c|c|c}
    \hline
    \multirow{2}[4]{*}{\textbf{Attributes}} & \multicolumn{2}{c|}{\textbf{Environment Conditions}} & \multicolumn{3}{c|}{\textbf{RoadNetwork and Traffic Guidance}} & \multicolumn{5}{c}{\textbf{Dynamic Objects}} \bigstrut\\
\cline{2-11}          & \textbf{Weather} & \textbf{Light} & \textbf{LaneNum} & \textbf{SpeedLimit} & \textbf{CollisionLocation} & \textbf{DrivingActions} & \textbf{CrashType} & \textbf{DrivingDirections} & \textbf{RunningLanes} & \textbf{ParticipantsNumber} \bigstrut\\
    \hline
    \textbf{SoVAR} & 77.33\% & 93.33\% & 93.33\% & 100\% & 96.00\%  & 76.33\%  & 90.67\% & 91.66\% & 77.00\% & 100\% \bigstrut\\
    \hline
    \textbf{SoVAR\_N} & 69.33\% & 81.33\% & 76.00\%  & 91.33\% & 95.33\% & 55.00\% & 90.00\% & 91.33\% & 73.00\%  & 100\% \bigstrut\\
    \hline
    \textbf{AC3R} & 54.67\% & 76.00\% & 70.67\% & 82.67\% & 78.67\% & 43.67\% & 11.67\% & 39.30\% & 25.33\% & 94.67\% \bigstrut\\
    \hline
    \end{tabular}%
  \label{tab:extract information acc}%
  \vspace{-5pt}
\end{table*}%

\textbf{Test generation.} The simulations generated by SoVAR contain the essential information needed to reconstruct the car crashes described in the accident reports. However, these simulated crashes do not meet the specifications of the test cases because they lack proper test oracles and do not permit any exogenous interference. Specifically, the generated simulations cannot verify if the behavior of the ego car is acceptable, and the ego car cannot freely interact with the simulated environment. To address this, SoVAR automatically derives system-level test cases from the output of the trajectory planning module, allowing the ego car to choose a trajectory different from the one described in the accident report to avoid collisions. For the simulation scenarios generated based on the reports, SoVAR creates multiple test scenarios by designating each traffic participant as the ego vehicle. The starting point of each ego vehicle is set to the initial point of the generated waypoints, while the NPC vehicles follow the waypoints provided by the SoVAR trajectory planning module. Collision scenarios involving ego vehicles are recorded during these runs. However, some collisions are not caused by the ego vehicle, such as when an NPC vehicle crashes into a legally parked ego vehicle. To minimize false positives, SoVAR counts collision scenarios in which the ego vehicle's speed is not near zero.

\section{Experimental Design}

In this section, we introduce the experimental design, including the experiment settings, evaluation metric and baseline in the experiments. 
All experiments are performed on Ubuntu 21.10 desktop with GeForce RTX 4070, one 16-core processor at 3.80GHz, and 32GB RAM.

SoVAR aims to use accident reports as information sources to reconstruct accident scenarios and convert them into test cases for evaluating ADS. To this end, we empirically explore the following three research questions (RQ):

\begin{itemize}[leftmargin=*]

\item RQ1: How effective is information extraction from accident reports with SoVAR?

\item RQ2: What is the effectiveness of SoVAR in producing generalizable simulations with the intended impact?

\item RQ3: Do tests derived from generalizable scenarios find crashes in autonomous driving systems?
\end{itemize}

\subsection{Experiment Settings}

\textbf{Dataset and LLM.} To investigate the research questions, we acquired data from the National Motor Vehicle Crash Causation Survey database, maintained by the National Highway Traffic Safety Administration (NHTSA)~\cite{NHSTA}. The NHTSA adheres to the Model Minimum Uniform Crash Criteria (MMUCC) guidelines~\cite{NHSTAguideline}, which outline the primary factors contributing to crashes. To obtain the necessary accident reports, we divide the NHTSA police reports according to crash type and then randomly select a sample of reports from each category. We obtained NHTSA reports, enabling comprehensive coverage of diverse environmental conditions such as daytime and nighttime scenarios, various road types, as well as different crash types. We adopted GPT-4~\cite{achiam2023gpt} as the information extraction model. We chose GPT-4 because it is the current state-of-the-art LLM, widely known and easily accessible.


\textbf{Simulation Platform.} We connected the scene recovery outcomes with the leading virtual testing platform, LGSVL, widely used in academia and industry. LGSVL simulator is a high-fidelity simulation tool specifically designed for autonomous driving~\cite{DBLP:conf/itsc/RongSTLLMBUGMAK20}. Its advanced engine enables comprehensive end-to-end simulation and seamless integration with Apollo, allowing numerous virtual testing scenarios to be generated. In our study, we implemented SoVAR in the simulation environment built with Baidu Apollo 6.0~\cite{apollo} (the ADS under test) and SORA-SVL simulator~\cite{SORA}.

\textbf{Constraints Solver.} 
We employ Z3~\cite{Z3}, an efficient Satisfiability Modulo Theories (SMT) solver, to compute the trajectories of participants. Z3 is renowned for its efficiency and incorporates state-of-the-art algorithms that allow it to tackle large-scale problems swiftly and accurately. Z3 has garnered widespread adoption in both academia and industry for various applications, including software analysis, cyber-physical systems, and beyond.

\subsection{Evaluation Metric and Baseline Comparison}\label{section:metric}

\noindent\textbf{\emph{Metric.}} 
To evaluate the success rate of accident scenario reconstruction, we propose using the "Scenario Reconstruction Rate" metric to measure the effectiveness of SoVAR in reconstructing scenarios from accident reports. Here an accident that is successfully reconstructed if the participants move along the generated trajectories can replay the same collision. And we express it as follows:

\begin{equation}
SRR= \frac{1}{N_{r}}\sum_{i=1}^{N_{r}}\mathbbm{1} \left[ \bigwedge_{\forall tr\in TR_i} SIM(tr, \mathbb{MAP})==1 \right]
\end{equation}

\noindent where $N_r$ represents the total number of accident reports, $\bigwedge$ means logical AND, $TR_i$ represents the set of accident-related trajectories generated by the $i$-th report, $\mathbb{MAP}$ represents the map used to reconstruct the scenario, and $SIM$ is a function that evaluates whether the trajectory, derived from correctly parsed collision actions, is correct when executed in the simulator. Specifically, $SIM$ outputs 1 if the generated trajectory does not illegally cross any lines and the collision angle matches the report description; otherwise, it outputs 0. Additionally, $\mathbbm{1}$ is an indicator function mapping  
a boolean condition to a value in \{0, 1\}: If the condition is true, 
it returns 1; otherwise, it returns 0.


\noindent\textbf{\emph{Baseline.}} 
To evaluate ADS, there are various research works focusing on scenario reconstruction using accident information. Some methods leverage video recordings~\cite{DBLP:conf/icra/BashettyAF20,DBLP:conf/issta/Zhang023} and accident sketches~\cite{DBLP:conf/aitest/GambiNAF22} as information sources to directly identify and extract trajectories of traffic participants. In contrast, our approach, similar to AC3R~\cite{DBLP:conf/sigsoft/GambiHF19}, automatically plans and generates trajectories based on the textual description provided in accident reports.

To assess the effectiveness of SoVAR in accident information extraction and its capability to generate generalized scenarios, we employ AC3R as a baseline for comparison. AC3R integrates NLP techniques with a domain-specific ontology to extract pertinent information from accident reports and subsequently generates trajectories using hardcoded rules with fixed patterns. It should be noted that the AC3R method requires accident reports to be segmented before inputting into the information extraction module. To demonstrate the best performance of AC3R, we segment all reports for evaluation. In contrast, to showcase SoVAR's ability to handle complex text, we use raw, unsegmented accident reports as input to SoVAR.
Additionally, to further validate the effectiveness of linguistic pattern prompting in information extraction, we use SoVAR\_N as another baseline. The SoVAR\_N approach leverages GPT-4 to directly extract information from accident reports without employing any linguistic pattern prompting.

\section{Result Analysis and Discussion}

\subsection{Answer to RQ1}

To evaluate the effectiveness of information extraction from accident reports with SoVAR, we randomly select 150 reports encompassing diverse environmental conditions, road structures, and vehicle behaviors. Then, two of the authors independently analyze each accident report to establish the ground truth of extracted information. A third author is involved in a group discussion to resolve conflicts and reach agreements. Next, we execute the information extraction modules of SoVAR, SoVAR\_N, and AC3R, respectively, and automatically calculate the average extraction accuracy of each attribute across all selected reports.



\textbf{Results.}
Table~\ref{tab:extract information acc} illustrates the accuracy of various methods for extracting information from accident reports. The extraction accuracy of SoVAR for the environment conditions layer, road network layer, and dynamic object layer are 85.33\%, 96.44\%, and 87.13\%, respectively. Compared to the SoVAR\_N and AC3R methods, employing GPT-4 with linguistic patterns prompting can significantly enhance accuracy across all attribute information extractions. Specifically, the average accuracy of the SoVAR method surpassed that of the SoVAR\_N and AC3R methods by 7.3\% and 31.9\%, respectively. We observe that SoVAR and AC3R exhibit higher accuracy in extracting straightforward information such as SpeedLimit from the accident report. However, they demonstrate relatively lower effectiveness in extracting complex driving actions with contextual dependencies.

\begin{table}[]\scriptsize
  \centering
  \caption{The number of successfully reconstructed scenarios and scenario reconstruction rate under different road types.}
  \vspace{-10pt}
    \begin{tabular}{lllllllllll}
    \hline
    \multicolumn{2}{c|}{\multirow{2}[4]{*}{\textbf{Type}}} & \multicolumn{3}{c|}{\textbf{Intersection}} & \multicolumn{3}{c|}{\textbf{T-junction}} & \multicolumn{3}{c}{\textbf{Straight Road}} \bigstrut\\
\cline{3-11}    \multicolumn{2}{c|}{} & \multicolumn{1}{c|}{\textbf{I1}} & \multicolumn{1}{c|}{\textbf{I2}} & \multicolumn{1}{c|}{\textbf{I3}} & \multicolumn{1}{c|}{\textbf{T1}} & \multicolumn{1}{c|}{\textbf{T2}} & \multicolumn{1}{c|}{\textbf{T3}} & \multicolumn{1}{c|}{\textbf{S1}} & \multicolumn{1}{c|}{\textbf{S2}} & \multicolumn{1}{c}{\textbf{S3}} \bigstrut\\
    \hline
    \multicolumn{1}{c|}{\multirow{2}[4]{*}{\textbf{SoVAR}}} & \multicolumn{1}{c|}{\textbf{Num.}} & \multicolumn{1}{c|}{47} & \multicolumn{1}{c|}{45} & \multicolumn{1}{c|}{48} & \multicolumn{1}{c|}{35} & \multicolumn{1}{c|}{36} & \multicolumn{1}{c|}{38} & \multicolumn{1}{c|}{42} & \multicolumn{1}{c|}{41} & \multicolumn{1}{c}{40} \bigstrut\\
\cline{2-11}    \multicolumn{1}{c|}{} & \multicolumn{1}{c|}{\textbf{SRR}} & \multicolumn{1}{c|}{94\%} & \multicolumn{1}{c|}{90\%} & \multicolumn{1}{c|}{96\%} & \multicolumn{1}{c|}{70\%} & \multicolumn{1}{c|}{72\%} & \multicolumn{1}{c|}{76\%} & \multicolumn{1}{c|}{84\%} & \multicolumn{1}{c|}{82\%} & \multicolumn{1}{c}{80\%} \bigstrut\\
    \hline
    \multicolumn{11}{l}{\textbf{*Note that the scenario reconstruction rate for AC3R is 0.}} \bigstrut[t]\\
    \end{tabular}%
  \label{tab:4}%
  \vspace{-5pt}
\end{table}%

\begin{table*}[htbp]\scriptsize
  \centering
  \caption{Fault localization analysis of not successfully reconstructed scenarios.}
  \vspace{-10pt}
    \begin{tabular}{c|c|c|c|c|c|c|c|c|c|c|c|c|c|c|c|c}
    \hline
    \multirow{2}[4]{*}{\textbf{Method}} & \multicolumn{4}{c|}{\textbf{Trajectory Planning}} & \multicolumn{4}{c|}{\textbf{Simulation Crash Type}} & \multicolumn{4}{c|}{\textbf{Simulation Crossing}} & \multicolumn{4}{c}{\textbf{Total}} \bigstrut\\
\cline{2-17}          & \textbf{I} & \textbf{T} & \textbf{S} & \textbf{Percentage} & \textbf{I} & \textbf{T} & \textbf{S} & \textbf{Percentage} & \textbf{I} & \textbf{T} & \textbf{S} & \textbf{Percentage} & \textbf{I} & \textbf{T} & \textbf{S} & \textbf{Percentage} \bigstrut\\
    \hline
    \textbf{SoVAR} & 0.00\% & 0.00\% & 1.33\% & 1.33\% & 1.78\% & 8.00\% & 4.22\% & 14.00\% & 0.44\% & 1.11\% & 0.44\% & 2.00\% & 2.22\% & 9.11\% & 6.00\% & 17.33\% \bigstrut\\
    \hline
    \textbf{AC3R} & 21.62\% & 24.32\% & 22.52\% & 68.47\% & 2.70\% & 2.70\% & 3.60\% & 9.01\% & ---     & ---     & ---     & ---     & 24.32\% & 27.03\% & 26.13\% & 77.48\% \bigstrut\\
    \hline
    \end{tabular}%
  \label{tab:5}%
  \vspace{-5pt}
\end{table*}%

\begin{table}[]\small
  \centering
  \caption{The number of collisions caused by ADS in generated test scenarios using different methods.}
  \vspace{-10pt}
    \begin{tabular}{c|c|c|c|c}
    \hline
    \textbf{Collision} & \textbf{Intersection} & \textbf{T-junction} & \textbf{Straight Road} & \textbf{Total} \bigstrut\\
    \hline
    \textbf{SoVAR} &   15    &    4   &   6    & 25 \bigstrut\\
    \hline
    \textbf{Random} &   1    &   1   &    2   &  4 \bigstrut\\
    \hline
    \end{tabular}%
  \label{tab:6}%
  \vspace{-10pt}
\end{table}%

\textbf{Discussion.}
According to the experimental analysis presented in Table~\ref{tab:extract information acc}, it is evident that utilizing LLMs, such as GPT-4, to extract information from accident reports is a highly effective method. Additionally, employing well-designed linguistic patterns for prompting can significantly enhance the LLM's ability to extract information. While our method achieves good extraction results, there is still room for improvement in the accuracy of attributes such as driving actions before a collision. Upon reviewing reports containing extraction errors, we identified that many inaccuracies stem from implicit actions described within accident report descriptions. For instance, a retrograde action might be described simply as "crossing the double yellow line." Additionally, there is no consistent pattern for how each accident report from the NHTSA database is structured. By applying a large language model with stronger generalization and inference capabilities, or by fine-tuning the model after manually labeling the accident information, the information extraction effect of SoVAR can be further optimized.


\begin{figure*}[htbp]
	\centering
    \includegraphics[width=\linewidth, height=0.25\linewidth]{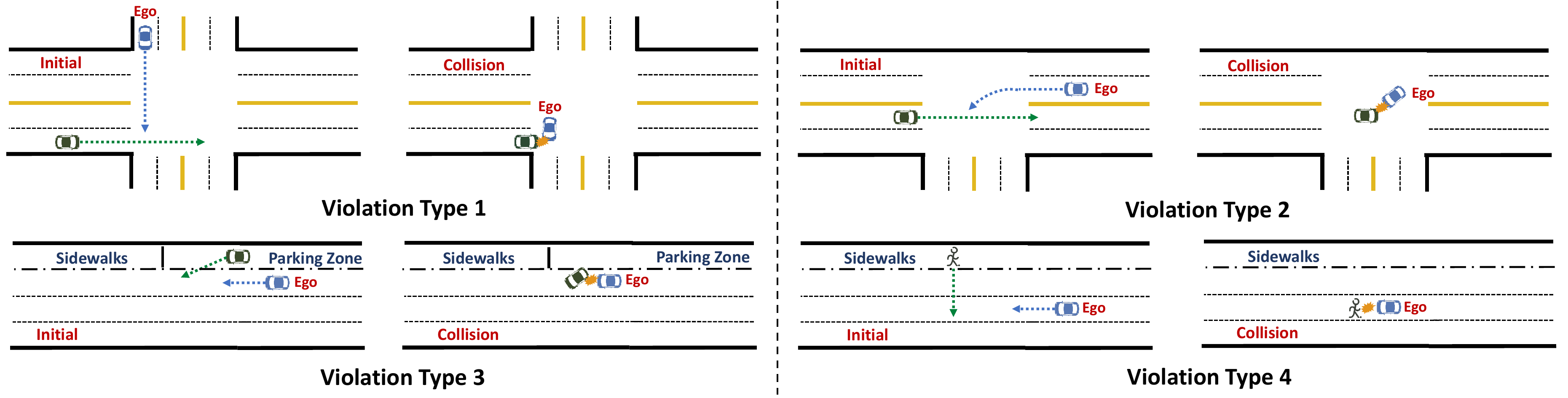}%
    \vspace{-10pt}
	\caption{Visualization samples showing the safety violations of Apollo detected by SoVAR on different road types.}
	\label{fig.4}
\end{figure*}

\subsection{Answer to RQ2}

To ascertain the generalizability of the SoVAR method, i.e., its capacity to replicate accidents across roads of varying lengths and widths depicted on maps, we reconstruct scenarios based on accident location reports from the NHTSA database. These reports encompass accidents occurring on straight roads, T-junctions, and intersection locations. Specifically, we randomly select 50 accident reports for each road type for experimental purposes, and SoVAR is executed three times for each report, ensuring that each run covers different roads on the San Francisco map. We tally the number of successful reconstruction scenarios for each type of road accident. Additionally, for scenarios that are not fully successfully reconstructed, we analyze the causes and classify them into three categories: trajectory planning program crashes, crash type mismatches during simulation generation, and instances of driving actions crossing the line during simulation generation.

\textbf{Results.}
Table~\ref{tab:4} shows the number of successfully reconstructed scenarios and SRR metrics under different road types. The average scenario reconstruction rate (SRR) of accidents occurring at intersections, T-junction roads, and straight roads are 93.3\%, 72.7\%, and 82.0\% respectively. Since the length and width of all roads chosen for reproduction in the experiment are inconsistent with those of AC3R, the experiments on AC3R found that none of the scenarios can be successfully reconstructed for simulation testing.
To further analyze the recurrence results, Table~\ref{tab:5} presents the fault localization analysis of scenarios during their reconstruction.
The proportion of scenarios generated by SoVAR that cannot be completely reconstructed is 17.33\%. Most of these scenarios fail due to Crash Type simulation errors, accounting for 80.8\% of all failed reconstruction attempts.
Since AC3R does not generate adaptive waypoints according to lane width and length during operation and lacks a mechanism to map waypoints to a map, we don't collect statistics on AC3R vehicles crossing the line, as shown in Table~\ref{tab:5}.  It can be seen from the table that SoVAR has a 60\% lower average accident reconstruction failure rate compared to AC3R. These results demonstrate that, in comparison to AC3R, SoVAR is more capable of generating generalized scenarios.


\textbf{Discussion.} Experiments demonstrate that SoVAR can generate generalized scenarios on lanes of varying lengths and widths. It is noted that 2\% of the scenarios reconstructed by SoVAR involve crossing-the-line situations. This issue arises due to slight errors in the lane width of the same road when the map creators manually construct the experimental simulation map. Most of the errors in the AC3R method stem from the trajectory planning module. Besides, the method for generating waypoints in AC3R is hard-coded. Given the complex combinatorial logic between adjacent actions, it is inevitable that some waypoints cannot be generated due to insufficient consideration of the code logic during execution.


\subsection{Answer to RQ3}

Here, we assess the effectiveness of test cases derived from simulations generated by SoVAR in detecting crashes involving ADS and identifying safety concerns. We randomly selected 200 accident reports from the NHTSA database for scenario reconstruction, of which 39.5\%, 26.5\%, and 34\% occurred at intersections, T-junctions, and straight roads, respectively. We then convert the reconstructed scenarios into test cases and evaluate the performance of the Apollo ADS. We compare SoVAR with a random scenario generation method. Specifically, the random scenario generation method utilizes the motion task of the ego vehicle generated by SoVAR and randomly creates reasonable NPC trajectories. We then separately quantify the number of collisions experienced by autonomous vehicles using both SoVAR and the random method during simulated test scenarios. Additionally, to further analyze the cause of the collisions, we capture the trajectories of the ego vehicle and the NPC vehicle when collisions occurred during test execution. Our thorough examination of ego and NPC actions aims to identify and classify various safety-violation scenarios that emerged.

\begin{figure}[]
	\centering
    \includegraphics[width=\linewidth, height=0.51\linewidth]{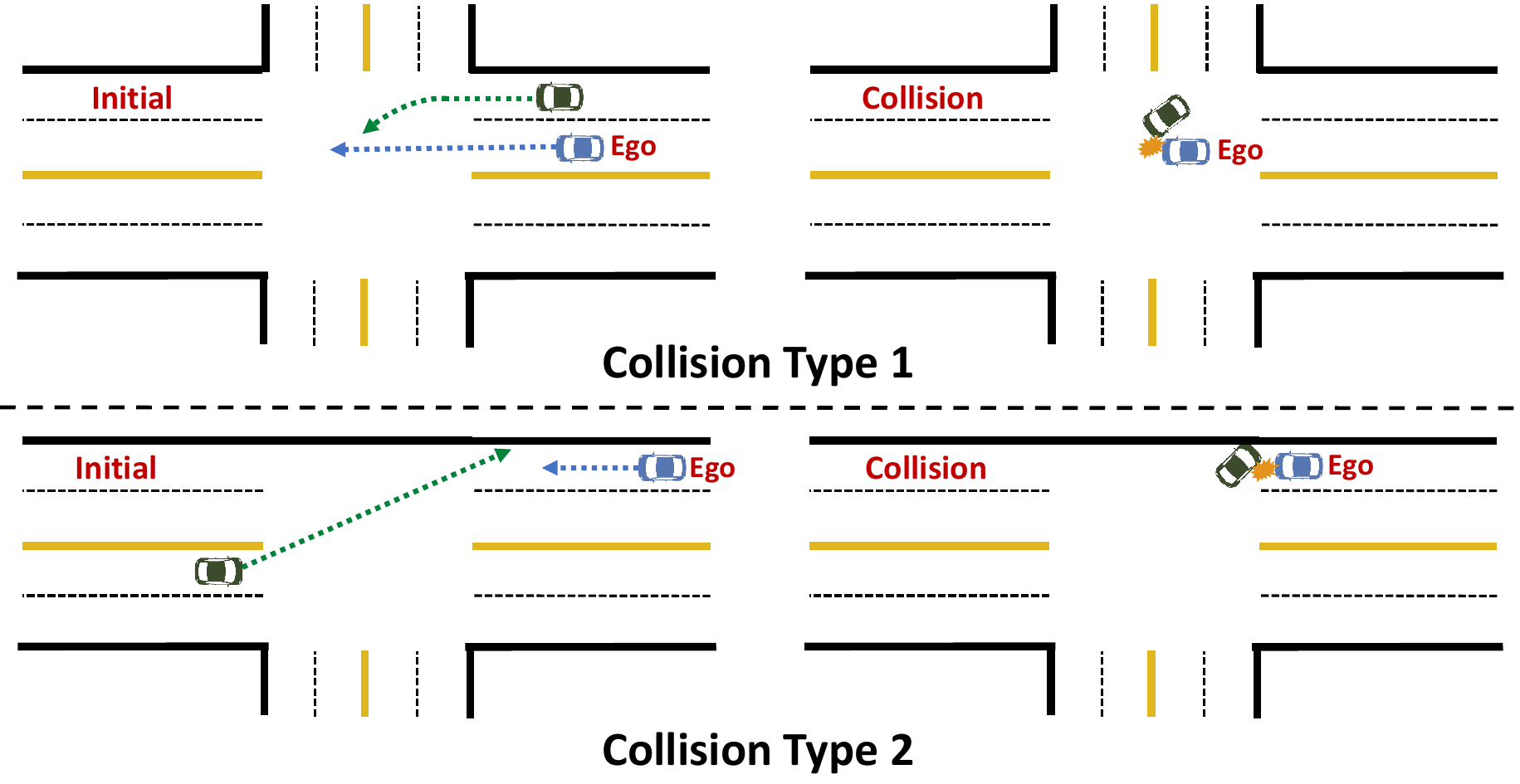}%
    \vspace{-10pt}
	\caption{Visualization samples showing another two collision types detected by SoVAR.}
	\label{fig.5}
\end{figure}

\textbf{Results.}  Table~\ref{tab:6} presents the number of collisions caused by self-driving cars in generated test scenarios using different methods. The table reveals that the number of collisions identified by SoVAR across various road types exceeds those found by the random method. Overall, the total number of collisions detected by SoVAR is more than six times higher than that detected by the random method. 
In our comprehensive analysis, we identify and summarize 5 types of safety violations in the Apollo ADS. Due to space limitations, we present four safety violation scenario types as shown in Figure~\ref{fig.4}. The left sub-figure of each group of safety violation types describes the location and driving route of the NPC and ego car in the initial scene, and the right sub-figure is a schematic diagram of the scene when a collision occurs. Illustrations for other types of safety violations can be found in the open-source repositories we provide. From the safety violation types illustrated in Figure~\ref{fig.4}, it is evident that various regular NPC actions, such as vehicle across, turn left, and drive into roads, can cause collisions involving the vehicle equipped with Apollo ADS. In addition, as shown in Figure~\ref{fig.5}, we summarize another two types of collisions, which are caused by irregular NPC actions.


\textbf{Discussion.}
Through an in-depth analysis of the results presented in Table~\ref{tab:6}, it becomes apparent that scene reconstruction utilizing information extracted from accident reports proves to be an effective method for testing the safety of ADS. This effectiveness stems from the fact that test scenarios generated from scenario descriptions provided in accident reports, which include driving actions among other details, are indeed crucial for ensuring safety. Furthermore, the safety violations we identify encompass erroneous behaviors in Apollo's operational capabilities. Specifically, violation type 1 demonstrates that when the NPC vehicle crosses an intersection at high speed, the ego vehicle fails to decelerate in time. Violation type 2 shows that when the NPC vehicle goes straight through an intersection, the ego vehicle turns without yielding, as required by traffic rules, resulting in a collision. Violation type 3 and violation type 4 indicate that the ego vehicle does not slow down in time when the NPC enters the road from the parking zone or sidewalks. The causes of the accidents in collision type 1 and collision type 2 are not attributed to Apollo. Collision type 1 shows that when an NPC turns left from the right lane at an intersection without traffic lights, the ego vehicle results in a collision with the NPC. Collision type 2 demonstrates that when an NPC crosses the double yellow line and drives in the opposite direction, the ego vehicle does not stop but continues to move forward, leading to a collision. Although in these cases, the NPCs violate traffic rules, we believe that ADS should have emergency avoidance capabilities when there is sufficient time to react. In such scenarios, the ADS's ability to issue early warnings and take countermeasures is crucial to ensuring safety.

\subsection{Threats to Validity}

\noindent\textbf{\emph{Data Selection.}} The selection of accident report data poses a primary threat to validity. We randomly selected a small number of police reports from the NHTSA database for experiments, because evaluating the effect of GPT-4 on extracting accident information requires time-consuming and laborious manual annotation. Nonetheless, the NHTSA database is a widely utilized dataset in academia, and we took measures to ensure that the selected reports encompass a range of environmental conditions, crash types, road geometries, and driving actions. We believe that our approach can be readily applied to other accident report datasets containing detailed descriptions of accident scenarios.

\noindent\textbf{\emph{Environmental Simulation.}} 
One of the validity threats arises from limitations in the environmental simulation within the simulator. the simulator lacks the simulation of LiDAR point cloud data during adverse weather conditions. To address these challenges, we directly send ground-truth perception data to Apollo in the form of messages to confirm that Apollo correctly perceives information, and then test downstream modules of perception in Apollo, including prediction, planning, and control~\footnote{https://github.com/ApolloAuto/apollo?tab=readme-ov-file\#architecture}. Actually, many ADS testing works use a similar simulation environment without considering the perception module~\cite{DBLP:conf/kbse/0008P00Y22,DBLP:conf/issta/ChengZX23,DBLP:conf/kbse/TangZ0WLW21}. Although we don't apply the impact of environmental factors in LGSVL, our experimental results demonstrate that our method remains effective in detecting flaws in Apollo. Moving forward, we plan to validate our approach on other simulators that support ADS perception capabilities.



\noindent\textbf{\emph{Count for Randomness.}} 
Another potential threat stems from the stochastic nature of constraint solving used in SoVAR, where Z3 is employed to generate scenarios that satisfy constraints. Due to the inherent multiple feasible solutions in constraint solving, the results of each run may exhibit slight variations. However, this characteristic also serves as an advantage for SoVAR, as it can produce diverse scenarios that meet the specified constraints. To mitigate this threat, we conducted each experiment three times to validate the method. Remarkably, the results of each SoVAR experiment consistently outperformed other baseline methods, underscoring the robustness and effectiveness of our approach.


\section{Related Work}


\subsection{Driving Accident Scenario Reconstruction}

Traffic accidents inherently encompass a vast amount of valuable semantic information, thereby facilitating the comprehension of the accident process by reconstructing the accident scenario~\cite{ali2021traffic,ling2024deep}. Recent research efforts have primarily concentrated on scenario reconstruction across various driving data sources, including leveraging sensor data collected during actual crashes~\cite{erbsmehl2009simulation}, textual descriptions~\cite{DBLP:conf/sigsoft/GambiHF19}, accident sketches~\cite{DBLP:conf/aitest/GambiNAF22}, and video recordings~\cite{DBLP:conf/icra/BashettyAF20,DBLP:conf/issta/Zhang023} to generate critical scenarios. Erbsmehl~\cite{erbsmehl2009simulation} recreates crashes by replaying the sensory data collected during actual crashes. However, this approach has limited applicability as it relies on naturalistic field operational data, which is not generally available. Gambi et al. implement an automatic crash construction tool AC3R, which automatically extracts information leveraging natural language processing and reproduces crashes in simulated environments~\cite{DBLP:conf/sigsoft/GambiHF19}. Subsequently, Gambi et al. present CRISCE, a semi-automated approach for generating simulations of critical scenarios from accident sketches that commonly complement crash reports~\cite{DBLP:conf/aitest/GambiNAF22}. Zhang et al. propose a panoptic segmentation model, M-CPS, designed to extract accident information from images or video recordings~\cite{DBLP:conf/issta/Zhang023}.



The key differences between the related work mentioned above and SoVAR are twofold: 
(1): Trajectories of traffic participants can be directly extracted from accident sketches or videos. Accident reconstruction from textual descriptions may result in multiple trajectories that satisfy the given text descriptions. (2): SoVAR can accurately extract accident information and generate generalized scenarios adaptable to different map structures.

\subsection{ADS Simulation Testing} 

Reconstructing core road scenarios from original accident information holds significant value for ADS simulation testing~\cite{DBLP:journals/tosem/TangZZZGLGLMXL23}. The reconstructed scenarios can serve as initial test scenarios for ADS fuzz testing~\cite{DBLP:conf/kbse/TianWYJ00LY22,DBLP:conf/ivs/ScheuerGA23}. Numerous studies have examined and unveiled crashes exhibited by ADS through the reconstruction of accident scenarios~\cite{DBLP:conf/sigsoft/GambiHF19,DBLP:conf/issta/Zhang023,DBLP:conf/ivs/ScheuerGA23,DBLP:conf/icra/BashettyAF20}. Gambi et al. present a crash construction tool, which generates test cases by incorporating test oracles to evaluate the performance of the DeepDriving self-driving car software based on simulations~\cite{DBLP:conf/sigsoft/GambiHF19}. Bashetty et al. adopt a framework that aims to extract 3D vehicle trajectories from dashcam videos. The framework then recreates these extracted scenarios to facilitate the testing of collision avoidance systems~\cite{DBLP:conf/icra/BashettyAF20}. Tian et al. extract behaviors from collision-related trajectories and perform behavior pattern mining to generate critical scenarios for testing Baidu Apollo~\cite{DBLP:conf/kbse/TianWYJ00LY22}. Zhang et al. propose a mutation algorithm based on the original accident scenario set to facilitate testing of Apollo ADS~\cite{DBLP:conf/issta/Zhang023}. Scheuer et al. implement the avoidable collision scenarios generation tool by extending focused collision descriptions using a multi-objective optimization algorithm~\cite{DBLP:conf/ivs/ScheuerGA23}.

\section{Conclusion}

This paper introduces and assesses SoVAR, a tool designed for automatically reconstructing crash scenarios from accident reports and testing ADS. SoVAR leverages LLM to extract detailed accident information, significantly enhancing LLM's text comprehension and parsing capabilities through the design of specialized linguistic patterns for extraction prompts. To enhance the method's generalization capability in reconstructing accident scenarios, SoVAR generates accident-related trajectories by solving a predefined set of trajectory specifications. Experimental results demonstrate that our method can successfully replicate accident reports on different map structures and the reconstructed simulation scenarios can identify various types of safety violations based on industrial-grade ADS. These findings underscore the crucial role of SoVAR in upholding the quality and dependability of ADS.



\bibliographystyle{ACM-Reference-Format}
\balance
\bibliography{reference}

\end{document}